\documentclass[preprint,amssymb,superscriptaddress,aps,prl,floatfix]{revtex4-1}
\usepackage{amsmath}
\usepackage{graphicx}
\usepackage{array}
\usepackage{dcolumn}
\usepackage{bm}
\usepackage[usenames,dvipsnames]{color}
\usepackage{epstopdf}
\usepackage{lipsum}

\usepackage{tcolorbox}
\usepackage{hyperref}
\hypersetup{
colorlinks = true,
urlcolor = blue,
citecolor= blue,
}

\usepackage{tikz}

\usepackage{siunitx}

\def\x{{\mbox{\boldmath$x$}}}

\newcommand{\bb}{\color{black} \normalfont}

\usepackage[normalem]{ulem}
\usepackage[textwidth=\dimexpr\textwidth-2cm\relax]{todonotes}
\makeatletter
\@mparswitchfalse%
\makeatother
\normalmarginpar 

\begin{document}

\preprint{APS/123-QED}

\title{Unifying theory of scaling in drop impact:\\ Forces \& maximum spreading diameter}

\author{Vatsal Sanjay}
 \email{vatsalsanjay@gmail.com}
\affiliation{	
	Physics of Fluids Group, Max Planck Center Twente for Complex Fluid Dynamics, and J. M. Burgers Center for Fluid Dynamics, University of Twente, P.O. Box 217, 7500AE Enschede, Netherlands
}
\author{Detlef Lohse}
\email{d.lohse@utwente.nl}
\affiliation{	
	Physics of Fluids Group, Max Planck Center Twente for Complex Fluid Dynamics, and J. M. Burgers Center for Fluid Dynamics, University of Twente, P.O. Box 217, 7500AE Enschede, Netherlands
}
\affiliation{
	Max Planck Institute for Dynamics and Self-Organisation, Am Fassberg 17, 37077 G{\"o}ttingen, Germany
}

\date{\today}


\begin{abstract}
The dynamics of drop impact on  a rigid surface 
-- omnipresent in nature and technology -- 
strongly depends on the droplet's velocity, its size, and 
its material properties. The main characteristics are the droplet's force exerted on the surface
and its maximal spreading radius. The crucial question is: How do they depend on the 
(dimensionless) control parameters, which are the Weber number $We$ (non-dimensionalized kinetic energy) and the Ohnesorge number $Oh$ (dimensionless viscosity)? Here we perform direct numerical simulations
over the huge parameter range $1\le We \le 10^3$ and $10^{-3}\le Oh \le 10^2$ and in particular develop a unifying theoretical approach, which is inspired by the Grossmann-Lohse theory
for wall-bounded turbulence [J. Fluid Mech. 407, 27 (2000); PRL 86, 3316 (2001)].
The key idea is to split the energy dissipation rate into the different phases 
of the impact process, in which different physical mechanisms  dominate. The theory can
consistently and quantitatively 
account for the $We$ and $Oh$ dependences of the maximal impact force and the maximal
spreading diameter over the huge parameter space.
It also clarifies why viscous dissipation plays a significant role during impact, even for
low-viscosity droplets (low $Oh$),  in contrast to what had been assumed in prior theories. 
	
\end{abstract}

\maketitle

\begin{figure}
	\centering
	\includegraphics[width=\linewidth]{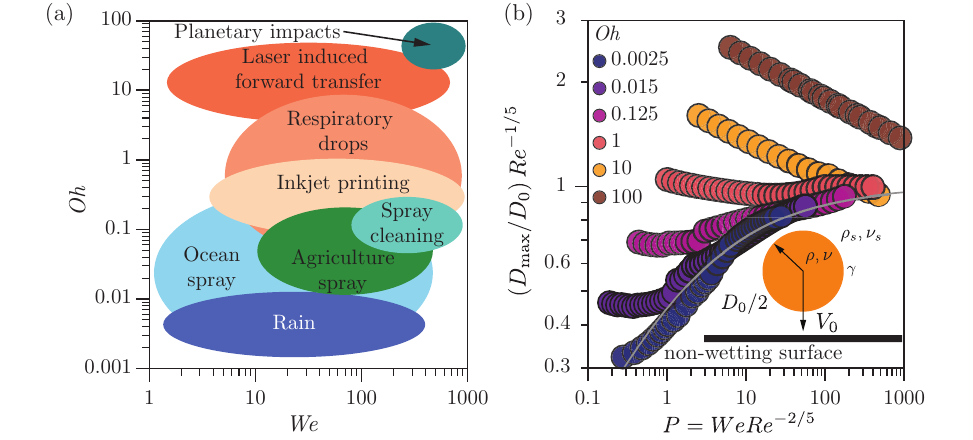}
	\caption{(a) Typical values for $We$ and $Oh$ of drop impact events with relevance in nature and technology.
 (b) Maximal droplet spreading diameter (compensated by  $Re^{1/5} $ for better visibility) vs the impact parameter $P = We Re^{-2/5}$ from our numerical simulations, for various
 values of $Oh$ (see legend). Also shown is the empirical fit proposed in
 ref.\ 
  \cite{laan2014maximum} (grey line), which  does not work for large 
 $Oh$ and has limitations for small $P$. 
 }
	\label{fig:intro}
\end{figure}

Droplet impact on a rigid surface is omnipresent and very relevant in 
nature and technology \cite{Josserand2016,yarin2006drop,Yarin2017}. 
Examples are 
rain  \cite{worthington1877xxviii}, inkjet printing  \cite{lohse2022fundamental}, spray coating \cite{kim2007spray}, criminal forensics \cite{smith2018influence},  
and respiratory droplets  \cite{bouroubia2021-arfm}, showing 
a plethora of different phenomena, depending on the droplet's  velocity $V_0$, 
its diameter $D_0$,
and its material properties (density $\rho$, dynamic  viscosity $\mu$,  and surface 
tension $\gamma$). In dimensionless form, these control parameters are 
commonly (and for Newtonian droplets according to the $\Pi$-theorem 
completely) 
expressed 
as Weber number $We\equiv \rho V_0^2 D_0 /\gamma $ and 
Ohnesorge number $Oh\equiv \mu / (\rho \gamma D_0 )^{1/2} $; the Reynolds number then follows
as $Re = \sqrt{We}/Oh$ and so does the so-called impact parameter as $P= We Re^{-2/5}$. 
The huge range over which the control parameters of the above natural and industrial
droplet impact events can occur
-- 
at least 5 orders of magnitude in $We$ and 4 order of magnitude in $Oh$
--
are visualized in the $We-Oh$ parameter space of
figure \ref{fig:intro}a. 
Gravity normally does not play any or hardly any role in  these impact processes and
in this paper it is assumed to be zero. We moreover assume
an axisymmetrical impact; for discussions on splashing, 
which sets in for very large impact velocities and at later
times during the impact event, 
we refer to \cite{sanjayzhang2022prl}.

At impact, the drop
encounters a normal reaction force \cite{Nearing1986,Mitchell2019,cheng2021drop}, which transforms its vertical momentum into radial spreading \cite{sanjayzhang2022prl}. During this spreading  phase, inertia drives the droplet outwards  until it reaches its  maximum diameter $D_\text{max} $ \cite{chandra1991collision}, where surface tension and viscosity collectively limit further deformation \cite{sanjay2024inertia}.
What is of particular relevance for applications 
are
this maximal  spreading diameter $D_\text{max}$ and the 
maximal normal force $F_\text{max}$ 
which the drop
exerts on the surface at impact. 
For the former,  traditionally,
scaling relations were sought for, such as
$D_\text{max} /D_0 \sim Re^{1/3}$ \cite{jorgensen2024deformation} or $\sim Re^{1/5}$ \cite{madejski1976solidification,fedorchenko2005effect} for viscous drop impact, $D_\text{max} /D_0 \sim We^{1/4}$ \bb for larger impact velocities 
\cite{Clanet2004}, 
or $D_\text{max} /D_0 \sim We^{1/2}$ for very large impact velocities 
\cite{bennett1993splat,Eggers2010,villermaux2011drop,Wildeman2016}. 
However, it is clear that none of these scaling relations can hold in 
the {\it  whole}  parameter space of figure \ref{fig:intro}a. Even empirically modelled  
transitions between two different 
scaling laws fail, as shown in figure 
\ref{fig:intro}b, where we compare our numerical simulations (as detailed below) with
the popular empirically  modelled smooth transition 
between two limiting scaling relations \cite{laan2014maximum}
\begin{equation}
    \frac{D_{\text{max}}}{D_0} = a_0 \frac{Re^{1/5 } P^{1/2}}{a_1 + P^{1/2}}.
    \label{laan}
\end{equation}
Here $a_0 = 1$ and $a_1 = 1.24 \pm 0.01$ are empirical constants obtained from fitting 
experimental data \cite{laan2014maximum}. 
While eq.\ (\ref{laan})  reasonably well describes the data for 
droplets with small viscosities 
(small $Oh$) and large impact velocities
(large $P$), it does so less for small $P$, and not at all for large $Oh$
(cf.\ fig.\ \ref{fig:intro}b).



\begin{figure}
\centering
\includegraphics[width=\textwidth]{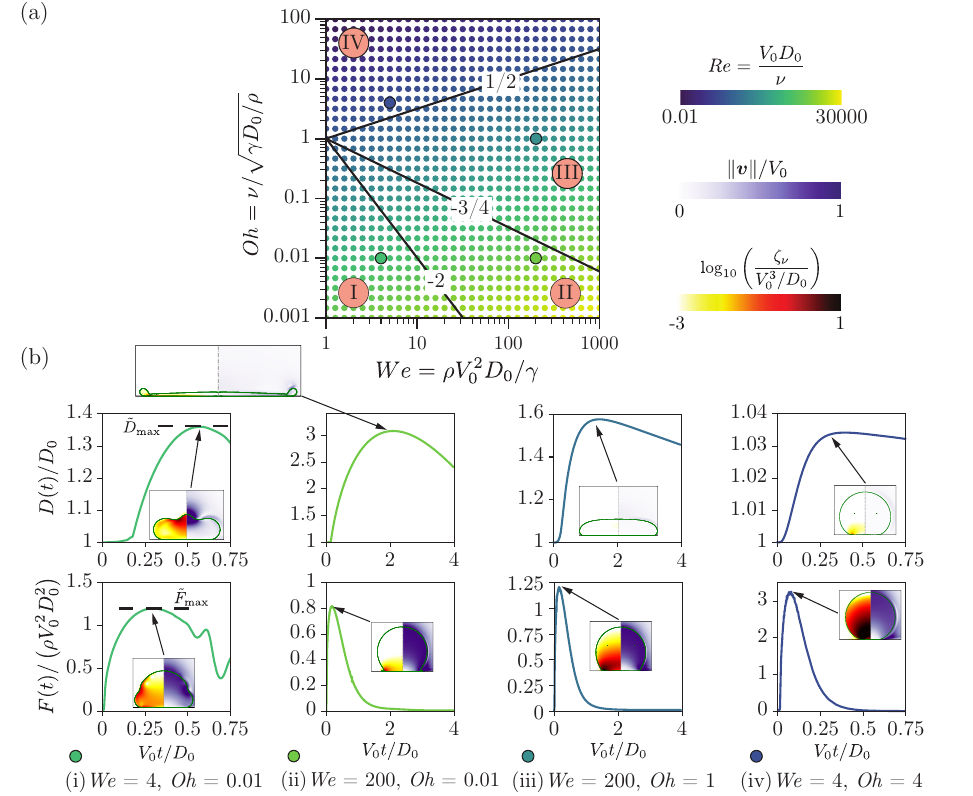}
\caption{(a) Phase space
 in the $Oh$-$We$ plane, illustrating the range of simulations conducted in this work, with simulation points (for clarity, only every 16$^{\text{th}}$ case is shown) colored according to the Reynolds number $Re = \sqrt{We}/Oh$.
 The three black lines (with slopes $-2$, $ -3/4$, 
 and $1/2$, respectively)
 delineate the  four different regimes of the drop spreading process
 (see table in fig.\ \ref{fig-table}): 
 Regime I: unbounded dissipation
 (small Oh and small We);
 Regime II: vertically bounded dissipation 
 (small Oh and large We);
 Regime III: fully bounded dissipation  
 (medium Oh and large We);
 Regime IV: no spreading phase 
 (large Oh). 
 (b) Four typical cases across the phase space, representing each of the four regimes. The plots show
 the spreading diameter $D(t)$ and the normal 
 reaction force $F(t)$. 
 We chose ($We, Oh$) = (i) ($4, 4$), (ii) ($4, 0.01$), (iii) ($200, 1$), and (iv) ($200, 0.01$). For each case, the insets show the drop at the moments of maximum impact force and later at maximum spreading diameter. The left part of each numerical snapshot depicts the dimensionless local viscous dissipation rates $\zeta_\nu (\x , t) $ \cite{supplMaterial} on a $\log_{10}$ scale, and the right part shows the local velocity field magnitude normalized by the impact velocity.}
\label{fig:1}
\end{figure}

\begin{figure}
\centering
\includegraphics[width=\linewidth]{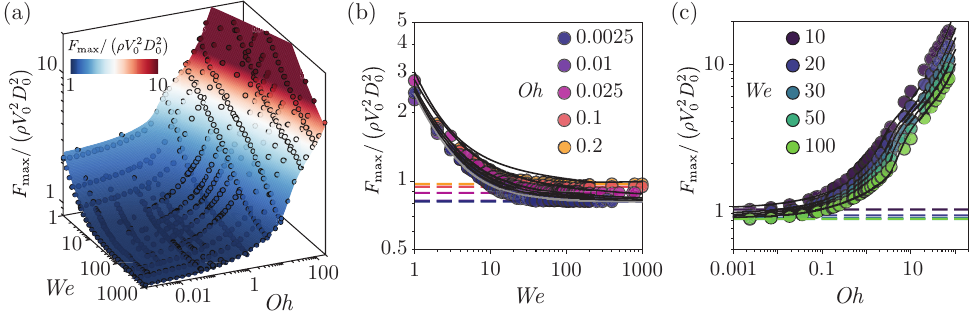}
\caption{Maximum impact force $F_{\text{max}}$ compensated with the inertial force scale $\rho V_0^2D_0^2$ as a function of (a) $Oh$ and $We$ in a 3D plot, (b) $We$ at different $Oh$ and (c) $Oh$ at different $We$.  The data points (only $\sim 4\%$ of them are shown) represent the simulation results, while the surface in (a) and the lines in (b)--(c) depict the results of the proposed theoretical model. The grey solid line in (b) represents the solution ignoring viscous dissipation. The dashed lines in (b) and (c) mark the asymptotes $We \gg 1$ and $Oh \ll 1$, respectively.}
\label{fig:2}
\end{figure}

\begin{figure}
	\centering
	\includegraphics[width=\linewidth]{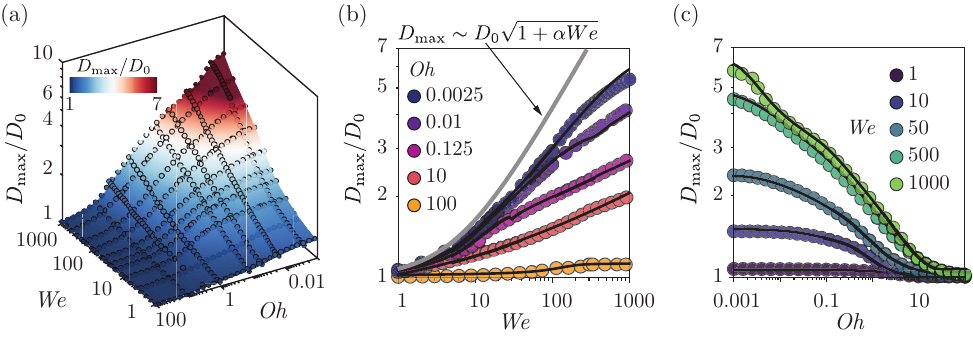}
	\caption{
 Maximum spreading diameter $D_{\text{max}}$ normalized by the initial drop diameter $D_0$ as a function of (a) $Oh$ and $We$ in a 3D plot, (b) $We$ for various $Oh$, and (c) $Oh$ for different $We$. The data points (only $\sim 4\%$ of them are shown)
 represent the simulation results, while the surface in (a) and the lines in (b)--(c) depict the results of the proposed theoretical model. The gray solid line in (b) represents the solution ignoring viscous dissipation.}
	\label{fig:3}
\end{figure}

The objective of this Letter is to achieve a unifying physical 
understanding of how the maximal spreading diameter $D_\text{max}$ and the maximal
normal force $F_\text{max}$ {\footnote{ 
We note  that in singular cases the 
first impact peak need not be the largest: As our recent studies \cite{sanjayzhang2022prl,sanjay2024inertia} have indicated, under 
very specific conditions ($Oh \sim \mathcal{O}(10^{-3})$, $We \approx 9$),  
a  second impact peak -- after droplet recoil and connected to 
the bouncing up of the droplet --  can exceed the first one due to a hydrodynamic singularity. For our considerations here it is not relevant.  }}
dependent on the system's control
parameters $We$ and $Oh$, for the whole 
huge relevant parameter space of fig.\ \ref{fig:intro}. To do so, we first 
perform over 16000
direct numerical simulations (with the volume-of-fluid solver Basilisk 
\cite{popinet-basilisk,vatsal_sanjay_2023_7598181}, detailed in the SI) 
of the drop impact process over this whole $We-Oh$ 
parameter space, see
fig.\ \ref{fig:1}a, and numerically obtain the dependencies
${F}_{\text{max}}\left(We , Oh\right)$ and 
${D}_{\text{max}}\left(We, Oh \right)$, see figures \ref{fig:2}
and \ref{fig:3}. 
We then, in the main part of this Letter, develop a unifying theory to 
account for these dependencies. It is inspired by 
Grossmann's and Lohse's unifying theory (``GL-theory'') 
for thermally driven turbulence
\cite{grossmann2000scaling, grossmann2001thermal, lohse2023ultimate,lohse2024}, 
whose key idea it is to spatially decompose  the energy dissipation rate
into boundary layer
and bulk contributions and to model these individually, 
based on the different flow physics in the boundary layer and in the bulk.
The  GL-theory
gives  the full dependencies of the response parameters (in that case the overall
heat transfer and turbulence intensity) on the control parameters. 
Here, for the droplet impact problem,
the decomposition of the energy dissipation rate will not 
be spatially, but temporally, namely 
splitting it into the impact phase and the spreading phase, 
each characterized by different scaling laws. 
This temporal decomposition allows us to disentangle 
the respective influences of these two phases 
on the maximal  force and the maximal spreading diameter, and to come to a 
unifying physical 
understanding of the dependencies 
${F}_{\text{max}}\left(We , Oh\right)$ and 
${D}_{\text{max}}\left(We, Oh \right)$ over the whole huge
control parameter 
space, consistent with our numerical results, cf.\ figures \ref{fig:2} and \ref{fig:3}.

{\it Energy balance  to obtain $D_\text{max}$:} We start by formulating the exact energy balance for the drop impact process \cite{Wildeman2016, sanjay_chantelot_lohse_2023},
\begin{align}
	\label{eq:energy}
	K_{\text{cm}}(t = 0) = K_{\text{cm}}(t) +  K_{\text{int}}(t) +  S(t) + mE_d(t),
\end{align}
 where $K_{\text{cm}}(t = 0) = mV_0^2/2$ is the center of mass kinetic energy at the moment of impact (fig.~\ref{fig:1}a), which is the total energy of the system with mass $m$. Upon impact and spreading, part of this initial kinetic energy remains as the center of mass kinetic energy $K_{\text{cm}}(t)$.
 Another part transforms into internal kinetic energy $K_{\text{int}}(t)$
 due to the redirection of vertical momentum into radial spreading. 
 The third term on the rhs is the gain $S(t)$ of surface energy through the 
 spreading process, 
 as compared to the drop's  surface energy at $t=0$, when it is minimal, because 
 the drop then is spherical. Finally, the last term $mE_d(t)$
 accounts for the total 
 viscous dissipation, which reduces the available energy over time, 
 where $E_d(t)$ is the viscous dissipation per unit mass up to time $t$. 
 It is obtained from  time-integration of the energy dissipation rate 
 per mass and time $\varepsilon (t)$ as $E_d(t) = \int_0^t \varepsilon (t) dt$.

 When is the maximal spreading $D_\text{max}$ achieved? 
 Scaling-wise, this spreading time 
$\tau_s$ is obtained from balancing inertia and capillarity, 
$\tau_s \sim  \sqrt{\rho D_0^3/\gamma}$, also reflecting the known 
 analogy between the first phase of drop spreading and the 
 oscillation of a free droplet
 \cite{Richard2002, Jha2020, sanjay_lohse_jalaal_2021}. 
The surface energy at that moment  $t=\tau_s$ of maximal spreading is directly related to the maximum spreading diameter $D_{\text{max}}$ as $S(t = \tau_s) \sim \gamma\left(D_{\text{max}}^2 - D_0^2\right)$. Moreover, 
at that moment,  
the drop's center-of-mass  kinetic energy and its 
internal kinetic energy are both zero, 
$ K_{\text{cm}} (\tau_s ) =  K_{\text{int}} (\tau_s ) = 0$,  as illustrated in the insets of fig.~\ref{fig:1}(b) \cite{afkhami2013numerical, sanjay_lohse_jalaal_2021}. 
Then the maximum spreading diameter 
simply results from balancing the initial kinetic energy with the surface energy increment and the total viscous dissipation during spreading ($0 \le  t \le  \tau_s$), 
\begin{align}
	\label{eq:GlobalDmax}
	K_{\text{cm}}(t = 0) = S(t = \tau_s) + mE_d(t = \tau_s).
\end{align}
We can rewrite eq.\ (\ref{eq:GlobalDmax}) as 
\begin{align}
\label{eq:GlobalDmax_v2}
	D_{\text{max}} &= D_0\sqrt{\left(\alpha_0 + \alpha_1We\left(1 - \frac{E_d(t = \tau_s)}{V_0^2}\right)\right)}. 
 \end{align}
 Here we have replaced the scaling relations with exact equalities, which  introduces the free parameters 
 $\alpha_i$,  $i=0, 1$. 

{\it Force balance to obtain $F_\text{max}$:}
To obtain the force balance, we differentiate  the energy transfer rates in eq.~\eqref{eq:energy} with respect to time. The derivative 
$\dot{K}_{\text{cm}}(t) = -F(t)V_{\text{cm}}(t)$ straightforwardly gives
the normal reaction force  $F(t)$, where  $V_{\text{cm}}(t)$ is the center of mass velocity.  We obtain the 
maximal normal  force $F_{\text{max}} \approx F (t= \tau_\rho )$
from using this  force balance at 
the inertial time scale $t = \tau_\rho \equiv D_0/V_0$,
\begin{align}
	\label{eq:GlobalFmax}
	F_{\text{max}} = \frac{1}{V_0}\left(\dot{K}_{\text{int}}(t = \tau_\rho) + \dot{S}(t = \tau_\rho) + m\varepsilon(t = \tau_\rho)\right). 
\end{align}
At this time $t= \tau_\rho$, the drop only deforms locally at the south pole, which contacts the substrate, while the north pole is still descending with velocity $V_0$ \cite{Eggers2010, sanjayzhang2022prl}. 
We emphasize that the maximum force occurs 
at $t = \tau_\rho$, much earlier than  
the maximum spreading at $t = \tau_s \gg \tau_\rho$ (fig.~\ref{fig:1}b). 

 Next, we  evaluate the internal kinetic energy of the spreading drop as $K_{\text{int}}(t) \sim mV_{\text{f}}^2/2$, where $V_{\text{f}}$ is the velocity at which the contact footprint $D_{\text{f}}$ grows on the rigid surface. Intriguingly, both the inertial \cite{wagner1932stoss, mandre2009precursors} and viscous \cite{hertz1881contact, langley2017impact, bilotto2023fluid, bertin2024similarity} regimes exhibit the same scaling behavior: $D_{\text{f}} \sim \sqrt{V_0D_0t}$. Consequently, $\dot{K}_{\text{int}}(t) \sim mV_0D_0/t^2$, see refs.\ 
 \cite{Philippi2016, supplMaterial} for a detailed discussion. To calculate the rate of surface energy change, we assume that  the drop behaves like a deformable cylinder of constant volume. Thus, the scales for the 
 time derivative 
 of internal energy and surface energy at the inertial time scale are given by
\begin{align}
	\dot{K}_{\text{int}}^* &\equiv \dot{K}_{\text{int}}(t = \tau_\rho) \sim \rho V_0^3D_0^2,\,\,\text{and}\\
	\dot{S}^* &\equiv \dot{S}(t = \tau_\rho) \sim \gamma D_0V_0,
\end{align}
respectively. We can then insert the energy and energy rate scales calculated above to simplify 
eq.\ \eqref{eq:GlobalFmax} to 
\begin{align}
	\label{eq:GlobalFmax_v2}
	F_{\text{max}} &= \rho V_0^2D_0^2\left(\beta_0 + \frac{\beta_1}{We} + \frac{\varepsilon(t = \tau_\rho)}{V_0^3/D_0}\right). 
\end{align}
Again we have replaced the scaling relations with exact equalities, which  introduces the free parameters  
$\beta_i$, $i=0,1$.

{\it Inertial limits $Oh \to 0$:}
Before we calculate the 
total viscous dissipation $E_d(t) $  at time 
$t = \tau_s$  in eq.\  \eqref{eq:GlobalDmax_v2} and the viscous dissipation
rate $\varepsilon (t) = \dot E_d (t)$
at time $t= \tau_\rho$ in eq.\ \eqref{eq:GlobalFmax_v2} in order to obtain 
$F_\text{max} (We, Oh) $
and $D_\text{max} (We , Oh)$
for general $We$ and $Oh$, we first discuss the inertial limits $Oh \to 0$  
of eqs.\ \eqref{eq:GlobalFmax_v2} and \eqref{eq:GlobalDmax_v2}. 
 The resulting expression $F_{\text{max}} = \rho V_0^2D_0^2\left(\beta_0 + \beta_1/We\right)$ 
 in this limit is identical to 
 the empirically relationship already proposed in ref.\ 
 \cite{sanjayzhang2022prl}, 
 which models the crossover from inertial  dominance (first term) to 
 capillary dominance (second term) with increasing $We$. 
This expression nicely approximates the impact force 
over two orders of magnitude in the Ohnesorge number $0.0025 < Oh < 0.2$, as shown in fig.\ref{fig:2} and \cite{sanjay2024inertia}. However, 
as expected, 
the agreement breaks down for moderate to large $Oh$ (fig.\ref{fig:2}(c)), 
reflecting the relevance of the viscous contribution.
Similarly, 
 the 
inertial limit of 
eq.\ \eqref{eq:GlobalDmax_v2} 
gives 
$D_{\text{max}}/D_0  \sim \sqrt{1 + \alpha We}$, which is shown in 
figure \ref{fig:3}b as grey line and clearly does not describe the data
for large $Oh$.

{\it Key idea to calculate the energy dissipation rates for general $(Oh, We)$:}
So it has become clear  that for the general case it is crucial to
calculate the energy dissipation rates
in eqs.\  \eqref{eq:GlobalDmax_v2} and
\eqref{eq:GlobalFmax_v2}.
To do so,   we 
got inspired by  the GL-theory \cite{grossmann2000scaling,grossmann2001thermal,grossmann2002prandtl,lohse2023ultimate} 
for wall-bounded turbulent thermal convection. 
The key idea of that theory is to spatially split the viscous dissipation rate 
into  boundary layer and  bulk contributions and estimate those separately,
reflecting the different flow physics in these two regions.
Instead of the spatial decomposition employed in that theory, here, 
we {\it temporally}  decompose the viscous dissipation rate, by
dividing the whole drop impact process into an impact phase (index ``$i$'')
and a spreading phase (index ``$s$'').

{\it Maximum impact force for general $We$ and $Oh$:} 
For the general case 
of  eq.\ (\ref{eq:GlobalFmax_v2}),   
we have to evaluate  the energy 
dissipation rate
$\varepsilon^* \equiv \varepsilon(t = \tau_\rho)$. 
During the impact phase, which takes place on
the inertial time scale $\tau_\rho$,  the drop's south pole halts \cite{renardy2003pyramidal, Biance2006}, creating a velocity gradient near the surface across a boundary layer with thickness 
$\lambda_i(t)$, observable in the black region of fig.~\ref{fig:1}(b). This gradient $V_0/\lambda_i(t)$ propagates to the north pole, building up a  viscous velocity gradient in 
the drop. The thickness of this boundary layer during the impact phase 
scales as  $\lambda_i(t) \sim \sqrt{\nu t}$, 
according to the Prandtl-Blasius boundary layer theory \cite{Prandtl1905, Blasius1908,Schlichting1979}. 
Here $\nu = \mu /\rho$ is the kinematic viscosity. 
In this phase
the viscous dissipation rate $\varepsilon_{i, \text{PB}}(t)$ within the volume $\Omega_{\nu, i}(t) \sim D_{\text{f}}(t)^2\lambda_i(t)$
is approximated as 
\begin{align}
	\label{eq:PB}
	\varepsilon_{i,\text{PB}}(t) \sim \frac{\nu}{D_0^3}\left(\frac{V_0}{\lambda_i(t)}\right)^2D_{\text{f}}(t)^2\lambda_i(t) \sim \sqrt{\frac{\nu t}{D_0^2}}\frac{V_0^3}{D_0}.
\end{align}

\begin{figure}
	\centering
	\includegraphics[width=\linewidth]{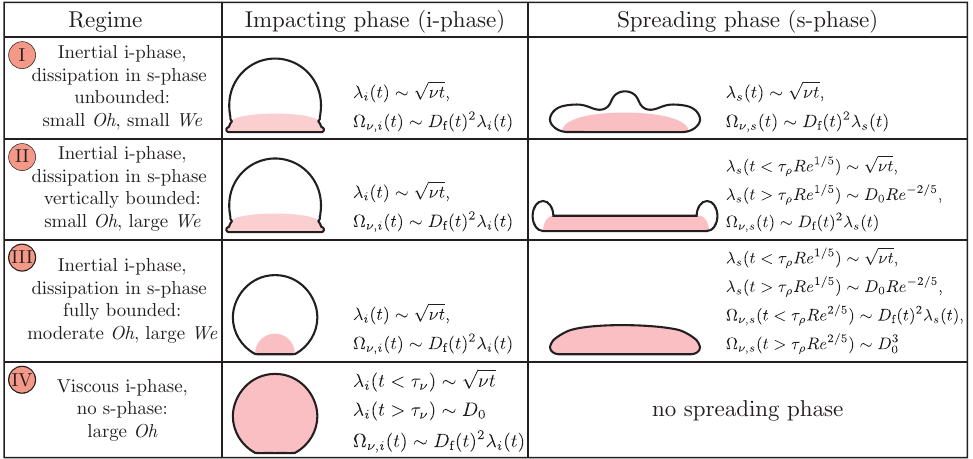}
	\caption{Sketches of the impacting phase (i-phase)
 and the spreading phase (s-phase) with the respective
 scaling relations for the four regimes I, II, III, and IV.  
 }
	\label{fig-table}
\end{figure}

For very viscous liquids (large $\nu$ and thus large 
$Oh$) the boundary layer thickness $\lambda_i \sim \sqrt{\nu t}$  very quickly reaches the full diameter $D_0$
of the droplet and then obviously no longer increases, $\lambda_i \sim D_0$, as illustrated in the last line of 
the table
in figure \ref{fig-table}. 
Then the energy dissipation rate $\varepsilon_{i,\text{PB}}(t)$
of eq.\ (\ref{eq:PB}), which holds during the Prandtl-Blasius 
phase of the impact phase, must be replaced by 
\begin{align}
	\label{eq:infty}
	\varepsilon_{i,\infty}(t) \sim \frac{\nu}{D_0^3}\left(\frac{V_0}{D_0 }\right)^2D_{\text{f}}(t)^2 D_0 \sim \left(\frac{\nu t}{D_0^2}\right)\frac{V_0^3}{D_0}.
\end{align}
Here we chose the index ``$\infty$" in analogy to the notation of the
GL-theory for thermally driven convection, where the corresponding regime, in which the boundary layer reaches the system size, is also noted with that index. 
This extra subphase  of the impacting phase can only occur when 
the 
viscous timescale $\tau_\nu \equiv D_0^2/\nu$
is faster than the impact timescale $\tau_\rho$,
i.e., when $\tau_\nu \ll \tau_\rho$, or, in other
words, when the drop impact 
  Reynolds number   $Re \equiv V_0D_0/\nu \ll   1$, i.e., indeed only in the
  viscous case. Then, to estimate the mean dissipation at time $t = \tau_\rho$, we must consider the two subphases of the impacting phase 
  separately.  

So,  in summary, the estimate for the dissipation  rate at
the time $t = \tau_\rho$ 
of  the maximal force $F_\text{max}$ is 
\begin{align}\label{fmax} 
\varepsilon^* \equiv \varepsilon(t = \tau_\rho) \sim 
\begin{cases}
    \overbrace{\frac{d}{dt}\left(\int_0^{\tau_\rho}\varepsilon_{i,\text{PB}}(t) dt\right)}^{\varepsilon_{i,\text{PB}}(t = \tau_\rho)}
    & \text{for } Re > 1\text{, and} \\[4ex]
    \underbrace{\frac{d}{dt}\left(\int_0^{\tau_\nu}\varepsilon_{i,\text{PB}}(t) dt\right)}_{\varepsilon_{i,\text{PB}}(t = \tau_\nu)} + \underbrace{\frac{d}{dt}\left(\int_{\tau_\nu}^{\tau_\rho}\varepsilon_{i,\infty}(t) dt\right)}_{\varepsilon_{i,\infty}(t = \tau_\rho) - \varepsilon_{i,\infty}(t = \tau_\nu)}
    & \text{for } Re < 1
\end{cases}
\end{align}
which, when filled into  eq.\ \eqref{eq:GlobalFmax_v2}, gives our final
result for the (nondimensionalized) maximal impact force, 
\begin{align}
	\label{eq:FinalFmax}
	\frac{F_{\text{max}}}{\rho V_0^2D_0^2} =\, \beta_0 + \frac{\beta_1}{We} + \left\{ \begin{aligned} &k_0 \left(Oh/\sqrt{We}\right)^{1/2} && \text{for}\,Re > 1\\ &m_0 + m_1\left(Oh/\sqrt{We} - 1\right) && \text{for}\,Re < 1 \end{aligned} \right.
\end{align}
Here, again, the scaling relations have been replaced by an equal sign and the corresponding prefactors 
(here $k_0$, $m_0$, $m_1$) 
that 
must be fitted to trustable data (see SI for details), 
in perfect analogy to what had to be done
in the GL-theory for thermal convection. 
The result is shown in  fig.~\ref{fig:2} and compared to the 
numerical data, which are excellently described.

{\it Maximum spreading diameter for general $We$ and $Oh$:} 
As seen above,  the viscous contribution $E_d(t=\tau_s)
  =\int_0^{\tau_s}   \varepsilon (t) dt  $  
in eq.~\eqref{eq:GlobalDmax_v2} is highly relevant 
to obtain the general case 
for the maximal spreading diameter $D_\text{max}$. 
To calculate it, as explained above, we have to take notice of the different
fluid dynamics in the impacting phase and in the spreading phase.
We correspondingly decompose the total energy dissipation into
\begin{align}
	\label{eq:dissipationIntegral}
	E_d(t = \tau_s) = \int_0^{\tau_\rho}\varepsilon_i(t)dt + \int_{\tau_\rho}^{\tau_s}\varepsilon_s(t)dt,
\end{align} 
where 
\begin{align}\label{eq14}
	\varepsilon_i(t) &\sim \frac{\nu}{D_0^3}\left(\frac{V_0}{\lambda_i(t)}\right)^2\Omega_{\nu,i}(t)\quad\text{and}\\
 \label{eq15}
	\varepsilon_s(t) &\sim \frac{\nu}{D_0^3}\left(\frac{V_\text{f}(t)}{\lambda_s(t)}\right)^2\Omega_{\nu,s}(t)
\end{align}
 are the viscous dissipation rates in the impact and spreading time intervals, respectively. 
 
The impact interval can be computed identically to what we have done already in equations~\eqref{eq:PB} and~\eqref{eq:infty} for the force calculation.  
What occurs in 
 the spreading phase depends on $Oh$ and $We$ and 
 can be classified into four different 
 regimes, see fig.~\ref{fig:1}a and the right column
 in the table in figure \ref{fig-table}:
\begin{enumerate}
    \item[I] Regime I (small $Oh$ and small $We$): 
    The viscous boundary layer remains engulfed inside the falling drop and never reaches the north pole during impact or spreading. In this regime, the viscous boundary layer $\lambda_s(t)$ develops due to the radial spreading of the drop's foot $D_{\text{f}}(t)$ on the rigid surface and follows the conventional Prandtl-Blasius scaling \cite{Prandtl1905, Blasius1908,Schlichting1979}. The volume $\Omega_{\nu, s}(t)$ where the dissipation occurs can be modeled as a cylinder with diameter $D_{\text{f}}(t)$ and height $\lambda_s(t)$.

    \item[II] Regime II (small $Oh$ and large $We$): The viscous boundary layer stays inside the deforming drop during the impact but reaches the north pole in the spreading phase, with the dissipation region confined within the drop.

    \item[III] Regime III (medium $Oh$ and large $We$): Similar to Regime II, but the dissipation region spreads throughout the drop during the spreading phase.

    \item[IV]  Regime IV (large $Oh$): The viscous boundary layer reaches the north pole during impact, causing dissipation throughout the drop. This regime is identical to the $\infty$-regime encountered while evaluating the viscous dissipation rate in eq.~\eqref{eq:infty}.
\end{enumerate}
We refer the readers to the supplementary material \cite{supplMaterial} for detailed calculations of the scaling behaviors in these four regimes and summarize them in the 
table in figure 
\ref{fig-table}. 

The transitions between these four 
regimes are characterized by the different physical balances resulting in specific scaling relationships (fig.\ \ref{fig:1}). 
The transition from Regime I to II occurs when the viscous boundary layer reaches the north pole during spreading. To determine the crossover time to this regime, we use the trajectory of the drop's north pole, given by $D_0^3/(V_0^2t^2)$, independent of $We$ and $Oh$ for inertial impacts \cite{Eggers2010}, and equate it to the growing boundary layer thickness $\lambda_s(t) \sim \sqrt{\nu t}$. Beyond this crossover, $\lambda_s(t)$ becomes time-invariant and equal to $D_0Re^{-2/5}$ \cite{Eggers2010}. Consequently, the transition from regime I to II occurs at $\tau_s \sim \tau_\rho Re^{1/5}$, corresponding to $Oh \sim We^{-2}$.
The transition from Regime II to III is marked by the dissipation region extending throughout the drop during spreading, occurring when $\Omega_{\nu, s}(t) \approx D_0^3$, giving $\tau_s \sim \tau_\rho Re^{2/5}$, corresponding to $Oh \sim We^{-3/4}$. Notably, once the dissipation has taken over the entire drop's volume, the drop's foot also becomes immobile, and the spreading phase ceases at $\tau_s \sim \tau_\rho Re^{2/5}$.
Finally, the transition from Regime III to IV takes place when the viscous boundary layer reaches the north pole during impact, corresponding to $\tau_\rho \sim \tau_\nu$, $Re \sim 1$, or $Oh \sim \sqrt{We}$, as already explained above.

The next  step is to plug in  the relevant timescales  in eq.\ (\ref{eq:dissipationIntegral}) and to 
evaluate the integrals with the estimates (\ref{eq14}) and (\ref{eq15}).  We obtain 
\begin{align}
	\label{eq:FinalDmax}
	 \frac{E_d(t = \tau_s)}{V_0^2} \sim \begin{cases}
		\left(a_0 + a_1We^{1/4}\right)/\sqrt{Re}\quad\text{in Regime I,}\\
		\left(b_0 + b_1Re^{1/10} + b_2Re^{-1/10}\sqrt{We} \right)/\sqrt{Re}\quad\text{in Regime II,}\\
		\left(c_0 + c_1Re^{1/10} + c_2Re^{3/10}\right)/\sqrt{Re}\quad\text{in Regime III,}\\
		d_0 + d_1Re + d_1/Re\quad\text{in Regime IV,}
	\end{cases}
\end{align}
which is then plugged into eq.\ (\ref{eq:GlobalDmax_v2}) to get the
final results for $D_\text{max} (We, Oh)$. 
Again, in eq.\  (\ref{eq:FinalDmax}),
the 
scaling relations have been replaced by prefactors that must be
fitted to trustable data of $D_\text{max} ( We, Oh )$, cf.\ SI.
The result is shown in fig.~\ref{fig:3} and compared to the numerical
data, which again are excellently described.

We emphasize that, remarkably,
even in the low-viscosity limit $Oh \to 0$ the 
viscous 
contribution  (\ref{eq:FinalDmax}) to the final result for $D_\text{max}$ 
(cf.\ eq.\ (\ref{eq:GlobalDmax_v2})) cannot be neglected and does 
contribute! This for examples holds for 
water-like systems (which have $Oh \ll 1$), 
as we had already realized in ref.\ 
\cite{sanjay_chantelot_lohse_2023}. This dissipation occurs at impact, as depicted in the insets of fig.~\ref{fig:1}. Even 
for $Oh \to 0^+$, this dissipation remains substantial.
We note that this also holds for the normal impact force (for $Re <1$),
see eq.\ (\ref{fmax}). 
Such a singular limit is 
akin to the dissipative anomaly observed in fully developed turbulence \cite{prandtl1904, onsager1949statistical, dubrulle2019beyond, eggers2018role}, and we have  been recently investigated similar
singular limits also 
in the context of  sheet retraction \cite{sanjay2022taylor}, sliding drops \cite{talukdar2024}, converging gravito-capillary waves \cite{kayal2024focusing}.

{\it Conclusions and outlook:} This study presents a comprehensive theoretical framework to elucidate the scaling laws governing the maximum impact force 
$F_\text{max} (We, Oh)$ 
and the maximal spreading diameter $D_\text{max} (We, Oh) $ of droplets upon impact. 
Inspired by the GL-theory 
for thermally driven turbulence, 
we systematically consider viscous dissipation rates 
across various regimes and allows for predictions. 
Our model covers five decades in the Ohnesorge number $Oh$ and three decades in the Weber number $We$, effectively mapping the transitions between the 
different regimes. 
Future research could  extend  
the present model in order to  incorporate 
the effects of gravity, which  influences 
the spreading diameter at moderate to high $Oh$ numbers.

Our results illustrate that viscous dissipation significantly affects the maximum impact force and spreading diameter, even in regimes typically considered inertial, i.e., when $Oh$ is small. 
For instance, the solid gray line in Fig.~\ref{fig:3}(b) represents the theoretical upper bound of the spreading diameter when viscous dissipation is neglected and totally fails to describe the data.  
Drop impact thus is an example for a singular limit in hydrodynamics. 
It underscores the importance of accounting for viscous effects in
hydrodynamics systems in general, 
even in low-viscosity systems, reflecting 
the persistent contribution of  
dissipation even  for $Oh \to 0^+$. 

Our results also demonstrate the capacity of the key idea of 
the GL-theory, namely to decompose the energy dissipation rate 
of a hydrodynamic systems, either locally as in the original GL-theory
\cite{grossmann2000scaling,grossmann2001thermal},
or here temporally, and suggests to try such an approach also to
other hydrodynamic problems.

\begin{acknowledgments}
{\it Acknowledgements:} The authors are grateful to Pierre Chantelot, Andrea Prosperetti, Uddalok Sen, and Vincent Bertin for the stimulating discussions, and 
to Anjali Suman for help with the implementation of the regression algorithm to find the prefactors in eqs.~\eqref{eq:FinalFmax} and~\eqref{eq:FinalDmax} \cite{supplMaterial}.
We also acknowledge financial support from NWO and from Canon. 
The numerical simulations were carried out on the national e-infrastructure of SURFsara, a subsidiary of SURF cooperation, the collaborative ICT organization for Dutch education and research. 
\end{acknowledgments}

\vspace{-5mm}

\bibliographystyle{prsty_withtitle}
\bibliography{refs}

\providecommand{\noopsort}[1]{}\providecommand{\singleletter}[1]{#1}%
\begin{thebibliography}{10}

\bibitem{Josserand2016}
C. Josserand and S.~T. Thoroddsen, {\em Drop impact on a solid surface}, Annu.
  Rev. Fluid Mech. {\bf 48},  365  (2016).

\bibitem{yarin2006drop}
A.~L. Yarin, {\em Drop impact dynamics: {S}plashing, {S}preading, {R}eceding,
  {B}ouncing...}, Annu. Rev. Fluid Mech. {\bf 38},  159  (2006).

\bibitem{Yarin2017}
A.~L. Yarin, I.~V. Roisman, and C. Tropea, {\em Collision {Phenomena} in
  {Liquids} and {Solids}} (Cambridge University Press, Cambridge, 2017).

\bibitem{worthington1877xxviii}
A.~M. Worthington, {\em {XXVIII}. {On} the forms assumed by drops of liquids
  falling vertically on a horizontal plate}, Proc. R. Soc. London, Ser. A {\bf
  25},  261  (1876).

\bibitem{lohse2022fundamental}
D. Lohse, {\em Fundamental fluid dynamics challenges in inkjet printing}, Annu.
  Rev. Fluid Mech. {\bf 54},  349  (2022).

\bibitem{kim2007spray}
J. Kim, {\em Spray cooling heat transfer: The state of the art}, Int. J. Heat
  Fluid Flow {\bf 28},  753  (2007).

\bibitem{smith2018influence}
F.~R. Smith, C. Nicloux, and D. Brutin, {\em Influence of the impact energy on
  the pattern of blood drip stains}, Phys. Rev. Fluids {\bf 3},  013601
  (2018).

\bibitem{bouroubia2021-arfm}
L. Bourouiba, {\em The Fluid Dynamics of Disease Transmission}, Annu. Rev.
  Fluid Mech. {\bf 53},  473  (2021).

\bibitem{sanjayzhang2022prl}
B. Zhang, V. Sanjay, S. Shi, Y. Zhao, C. Lv, and D. Lohse, {\em Impact forces
  of water drops falling on superhydrophobic surfaces}, Phys. Rev. Lett. {\bf
  129},  104501  (2022).

\bibitem{laan2014maximum}
N. Laan, K.~G. de~Bruin, D. Bartolo, C. Josserand, and D. Bonn, {\em Maximum
  diameter of impacting liquid droplets}, Phys. Rev. Appl. {\bf 2},  044018
  (2014).

\bibitem{Nearing1986}
M.~A. Nearing, J.~M. Bradford, and R.~D. Holtz, {\em Measurement of force vs.
  time relations for waterdrop impact}, Soil Sci. Soc. Am. J. {\bf 50},  1532
  (1986).

\bibitem{Mitchell2019}
B.~R. Mitchell, J.~C. Klewicki, Y.~P. Korkolis, and B.~L. Kinsey, {\em The
  transient force profile of low-speed droplet impact: measurements and model},
  J. Fluid Mech. {\bf 867},  300  (2019).

\bibitem{cheng2021drop}
X. Cheng, T.-P. Sun, and L. Gordillo, {\em Drop Impact Dynamics: Impact Force
  and Stress Distributions}, Annu. Rev. Fluid Mech. {\bf 54},  57  (2021).

\bibitem{chandra1991collision}
S. Chandra and C.~T. Avedisian, {\em On the collision of a droplet with a solid
  surface}, Proc. R. Soc. Lond. A {\bf 432},  13  (1991).

\bibitem{sanjay2024inertia}
V. Sanjay, B. Zhang, C. Lv, and D. Lohse, {\em The role of viscosity on drop
  impact forces}, arXiv preprint  (2024).

\bibitem{jorgensen2024deformation}
L. J{\o}rgensen, {\em Deformation of drops at low Reynolds number impact},
  Phys. Rev. Fluids {\bf 9},  083601  (2024).

\bibitem{madejski1976solidification}
J. Madejski, {\em Solidification of droplets on a cold surface}, Int. J. Heat
  Mass Transfer {\bf 19},  1009  (1976).

\bibitem{fedorchenko2005effect}
A.~I. Fedorchenko, A.~B. Wang, and Y.~H. Wang, {\em Effect of capillary and
  viscous forces on spreading of a liquid drop impinging on a solid surface},
  Phys. Fluids {\bf 17},    (2005).

\bibitem{Clanet2004}
C. Clanet, C. B{\'e}guin, D. Richard, and D. Qu{\'e}r{\'e}, {\em Maximal
  deformation of an impacting drop}, J. Fluid Mech. {\bf 517},  199  (2004).

\bibitem{bennett1993splat}
T. Bennett and D. Poulikakos, {\em Splat-quench solidification: estimating the
  maximum spreading of a droplet impacting a solid surface}, J. Mater. Sci.
  {\bf 28},  963  (1993).

\bibitem{Eggers2010}
J. Eggers, M.~A. Fontelos, C. Josserand, and S. Zaleski, {\em Drop dynamics
  after impact on a solid wall: {T}heory and simulations}, Phys. Fluids {\bf
  22},  062101  (2010).

\bibitem{villermaux2011drop}
E. Villermaux and B. Bossa, {\em Drop fragmentation on impact}, J. Fluid Mech.
  {\bf 668},  412  (2011).

\bibitem{Wildeman2016}
S. Wildeman, C.~W. Visser, C. Sun, and D. Lohse, {\em On the spreading of
  impacting drops}, J. Fluid Mech. {\bf 805},  636  (2016).

\bibitem{supplMaterial}
See Supplemental Material at ({URL} to be inserted by the publisher) for a
  description of numerical methods, details of the various regimes and
  transition., 2024.

\bibitem{Note1}
We note that in singular cases the first impact peak need not be the largest:
  As our recent studies \cite {sanjayzhang2022prl,sanjay2024inertia} have
  indicated, under very specific conditions ($Oh \sim \protect \mathcal
  {O}(10^{-3})$, $We \approx 9$), a second impact peak -- after droplet recoil
  and connected to the bouncing up of the droplet -- can exceed the first one
  due to a hydrodynamic singularity. For our considerations here it is not
  relevant.

\bibitem{popinet-basilisk}
S. Popinet and collaborators, Basilisk {C}: {V}olume of {F}luid method,
  \url{http://basilisk.fr}, 2013--now.

\bibitem{vatsal_sanjay_2023_7598181}
V. Sanjay, Code repository: {Impact} forces of water drops falling on
  superhydrophobic surfaces, \url{https://doi.org/10.5281/zenodo.7598181},
  2023.

\bibitem{grossmann2000scaling}
S. Grossmann and D. Lohse, {\em Scaling in thermal convection: a unifying
  theory}, J. Fluid Mech. {\bf 407},  27  (2000).

\bibitem{grossmann2001thermal}
S. Grossmann and D. Lohse, {\em Thermal convection for large Prandtl numbers},
  Phys. Rev. Lett. {\bf 86},  3316  (2001).

\bibitem{lohse2023ultimate}
D. Lohse and O. Shishkina, {\em Ultimate turbulent thermal convection}, Phys.
  Today {\bf 76},  26  (2023).

\bibitem{lohse2024}
D. Lohse and O. Shishkina, {\em {Ultimate Rayleigh--B\'enard turbulence}}, Rev.
  Mod. Phys. {\bf 96},  035001  (2024).

\bibitem{sanjay_chantelot_lohse_2023}
V. Sanjay, P. Chantelot, and D. Lohse, {\em When does an impacting drop stop
  bouncing?}, J. Fluid Mech. {\bf 958},  A26  (2023).

\bibitem{Richard2002}
D. Richard, C. Clanet, and D. Qu{\'e}r{\'e}, {\em Contact time of a bouncing
  drop}, Nature {\bf 417},  811  (2002).

\bibitem{Jha2020}
A. Jha, P. Chantelot, C. Clanet, and D. Qu{\'e}r{\'e}, {\em Viscous bouncing},
  Soft Matter {\bf 16},  7270  (2020).

\bibitem{sanjay_lohse_jalaal_2021}
V. Sanjay, D. Lohse, and M. Jalaal, {\em Bursting bubble in a viscoplastic
  medium}, J. Fluid Mech. {\bf 922},  A2  (2021).

\bibitem{afkhami2013numerical}
S. Afkhami and L. Kondic, {\em Numerical simulation of ejected molten metal
  nanoparticles liquified by laser irradiation: Interplay of geometry and
  dewetting}, Phys. Rev. Lett. {\bf 111},  034501  (2013).

\bibitem{wagner1932stoss}
H. Wagner, {\em {\"U}ber {S}to{\ss}-{\,}und {G}leitvorg{\"a}nge an der
  {O}berfl{\"a}che von {F}l{\"u}ssigkeiten}, Z. Angew. Math. Mech. {\bf 12},
  193  (1932).

\bibitem{mandre2009precursors}
S. Mandre, M. Mani, and M.~P. Brenner, {\em Precursors to splashing of liquid
  droplets on a solid surface}, Phys. Rev. Lett. {\bf 102},  134502  (2009).

\bibitem{hertz1881contact}
H. Hertz, {\em On the contact of elastic solids}, J. Reine Angew, Math. {\bf
  92},  156  (1881).

\bibitem{langley2017impact}
K. Langley, E.~Q. Li, and S.~T. Thoroddsen, {\em Impact of ultra-viscous drops:
  air-film gliding and extreme wetting}, J. Fluid Mech. {\bf 813},  647
  (2017).

\bibitem{bilotto2023fluid}
J. Bilotto, J.~M. Kolinski, B. Lecampion, J. Molinari, G. Subhash, and J.
  Garcia-Suarez, {\em Fluid-mediated impact of soft solids}, arXiv preprint
  (2023).

\bibitem{bertin2024similarity}
V. Bertin, {\em Similarity solutions in elastohydrodynamic bouncing}, J. Fluid
  Mech. {\bf 986},  A13  (2024).

\bibitem{Philippi2016}
J. Philippi, P.-Y. Lagr{\'e}e, and A. Antkowiak, {\em Drop impact on a solid
  surface: short-time self-similarity}, J. Fluid Mech. {\bf 795},  96  (2016).

\bibitem{grossmann2002prandtl}
S. Grossmann and D. Lohse, {\em {Prandtl} and {Rayleigh} number dependence of
  the Reynolds number in turbulent thermal convection}, Phys. Rev. E {\bf 66},
  016305  (2002).

\bibitem{renardy2003pyramidal}
Y. Renardy, S. Popinet, L. Duchemin, M. Renardy, S. Zaleski, C. Josserand,
  M.~A. Drumright-Clarke, D. Richard, C. Clanet, and D. Qu{\'e}r{\'e}, {\em
  Pyramidal and toroidal water drops after impact on a solid surface}, J. Fluid
  Mech. {\bf 484},  69  (2003).

\bibitem{Biance2006}
A.-L. Biance, F. Chevy, C. Clanet, G. Lagubeau, and D. Qu{\'e}r{\'e}, {\em On
  the elasticity of an inertial liquid shock}, J. Fluid Mech. {\bf 554},  47
  (2006).

\bibitem{Prandtl1905}
L. Prandtl,  in {\em Verhandlungen des III. Int. Math. Kongr., Heidelberg,
  1904} (Teubner, Leipzig, 1905), pp.\ 484--491.

\bibitem{Blasius1908}
H. Blasius, {\em {Grenzschichten in Fl\"ussigkeiten mit kleiner Reibung}}, Z.
  Math. Phys. {\bf 56},  1  (1908).

\bibitem{Schlichting1979}
H. Schlichting, {\em Boundary layer theory} (McGraw-Hill, New York, 1979).

\bibitem{prandtl1904}
L. Prandtl, {\em {\"U}ber Fl{\"u}ssigkeitsbewegung bei sehr kleiner Reibung},
  Math-Kongr, Heidelberg  484  (1904).

\bibitem{onsager1949statistical}
L. Onsager, {\em Statistical hydrodynamics}, Il Nuovo Cimento {\bf 6},  279
  (1949).

\bibitem{dubrulle2019beyond}
B. Dubrulle, {\em Beyond {Kolmogorov} cascades}, J. Fluid Mech. {\bf 867},  P1
  (2019).

\bibitem{eggers2018role}
J. Eggers, {\em Role of singularities in hydrodynamics}, Phys. Rev. Fluids {\bf
  3},  110503  (2018).

\bibitem{sanjay2022taylor}
V. Sanjay, U. Sen, P. Kant, and D. Lohse, {\em {Taylor-Culick} retractions and
  the influence of the surroundings}, J. Fluid Mech. {\bf 948},  A14  (2022).

\bibitem{talukdar2024}
J. Talukdar, U. Sen, C. Diddens, D. Lohse, and V. Sanjay, {\em Sliding drops on
  dry \& wet substrates}, {Working paper}  (2024).

\bibitem{kayal2024focusing}
L. Kayal, V. Sanjay, N. Yewale, A. Kumar, and R. Dasgupta, {\em Focusing of
  concentric free-surface waves}, arXiv preprint  (2024).

\end{thebibliography}


\providecommand{\noopsort}[1]{}\providecommand{\singleletter}[1]{#1}%
\begin{thebibliography}{10}

\bibitem{popinet-basilisk}
S. Popinet and collaborators, Basilisk {C}: {V}olume of {F}luid method,
  \url{http://basilisk.fr}, 2013--now.

\bibitem{prosperetti2009computational}
A. Prosperetti and G. Tryggvason, {\em Computational {Methods} for {Multiphase}
  {Flow}} (Cambridge university press, ADDRESS, 2009).

\bibitem{tryggvason2011direct}
G. Tryggvason, R. Scardovelli, and S. Zaleski, {\em {Direct} {Numerical}
  {Simulations} of {Gas}--{Liquid} {Multiphase} {Flows}} (Cambridge University
  Press, ADDRESS, 2011).

\bibitem{brackbill1992continuum}
J.~U. Brackbill, D.~B. Kothe, and C. Zemach, {\em A continuum method for
  modeling surface tension}, J. Comput. Phys. {\bf 100},  335  (1992).

\bibitem{landau2013course}
L.~D. Landau and E.~M. Lifshitz, {\em Fluid Mechanics -- Volume 6: Course of
  Theoretical Physics}, 2 ed. (Elsevier, ADDRESS, 1987).

\bibitem{ramirez2020lifting}
O. Ram{\'\i}rez-Soto, V. Sanjay, D. Lohse, J.~T. Pham, and D. Vollmer, {\em
  Lifting a sessile oil drop from a superamphiphobic surface with an impacting
  one}, Sci. Adv. {\bf 6},  eaba4330  (2020).

\bibitem{sanjay2023drop}
V. Sanjay, S. Lakshman, P. Chantelot, J.~H. Snoeijer, and D. Lohse, {\em Drop
  impact on viscous liquid films}, J. Fluid Mech. {\bf 958},  A25  (2023).

\bibitem{alventosa2023inertio}
L.~F.~L. Alventosa, R. Cimpeanu, and D.~M. Harris, {\em Inertio-capillary
  rebound of a droplet impacting a fluid bath}, J. Fluid Mech. {\bf 958},  A24
  (2023).

\bibitem{garcia2024skating}
P. Garc{\'\i}a-Geijo, G. Riboux, and J. Gordillo, {\em The skating of drops
  impacting over gas or vapour layers}, J. Fluid Mech. {\bf 980},  A35  (2024).

\bibitem{popinet-basilisk-momentum}
S. Popinet and collaborators, Basilisk {C}: conservative scheme for momentum
  advection, \url{http://basilisk.fr/src/navier-stokes/conserving.h},
  2013--now.

\bibitem{popinet-basilisk-viscosity}
S. Popinet and collaborators, Basilisk {C}: viscous stress solver library,
  \url{http://basilisk.fr/src/viscosity.h}, 2013--now.

\bibitem{popinet2018numerical}
S. Popinet, {\em Numerical models of surface tension}, Annu. Rev. Fluid Mech.
  {\bf 50},  49  (2018).

\bibitem{basiliskPopinet2}
S. Popinet and collaborators, Basilisk {C}: surface tension library,
  \url{http://basilisk.fr/src/tension.h}, 2013--2022.

\bibitem{popinet2009accurate}
S. Popinet, {\em An accurate adaptive solver for surface-tension-driven
  interfacial flows}, J. Comput. Phys. {\bf 228},  5838  (2009).

\bibitem{xu2005drop}
L. Xu, W.~W. Zhang, and S.~R. Nagel, {\em Drop splashing on a dry smooth
  surface}, Phys. Rev. Lett. {\bf 94},  184505  (2005).

\bibitem{Eggers2010}
J. Eggers, M.~A. Fontelos, C. Josserand, and S. Zaleski, {\em Drop dynamics
  after impact on a solid wall: {T}heory and simulations}, Phys. Fluids {\bf
  22},  062101  (2010).

\bibitem{Driscoll2011}
M.~M. Driscoll and S.~R. Nagel, {\em Ultrafast interference imaging of air in
  splashing dynamics}, Phys. Rev. Lett. {\bf 107},  154502  (2011).

\bibitem{riboux2014experiments}
G. Riboux and J.~M. Gordillo, {\em Experiments of drops impacting a smooth
  solid surface: {A} model of the critical impact speed for drop splashing},
  Phys. Rev. Lett. {\bf 113},  024507  (2014).

\bibitem{Josserand2016}
C. Josserand and S.~T. Thoroddsen, {\em Drop impact on a solid surface}, Annu.
  Rev. Fluid Mech. {\bf 48},  365  (2016).

\bibitem{sanjayzhang2022prl}
B. Zhang, V. Sanjay, S. Shi, Y. Zhao, C. Lv, and D. Lohse, {\em Impact forces
  of water drops falling on superhydrophobic surfaces}, Phys. Rev. Lett. {\bf
  129},  104501  (2022).

\bibitem{sanjay2024inertia}
V. Sanjay, B. Zhang, C. Lv, and D. Lohse, {\em The role of viscosity on drop
  impact forces}, arXiv preprint  (2024).

\bibitem{kolinski-2014-epl}
J.~M. Kolinski, L. Mahadevan, and S.~M. Rubinstein, {\em Drops can bounce from
  perfectly hydrophilic surfaces}, Europhys. Lett. {\bf 108},  24001  (2014).

\bibitem{sprittles2024gas}
J.~E. Sprittles, {\em Gas microfilms in droplet dynamics: When do drops
  bounce?}, Annu. Rev. Fluid Mech. {\bf 56},  91  (2024).

\bibitem{basiliskVatsal}
V. Sanjay, Code repository: {Impact} forces of water drops falling on
  superhydrophobic surfaces,
  \url{https://github.com/VatsalSy/Impact-forces-of-water-drops-falling-on-superhydrophobic-surfaces.git}
  (Last accessed: February 4, 2022), 2022.

\end{thebibliography}

\end{document}


\preprint{APS/123-QED}

\title{Supplementary material for: \\
	Unifying theory of scaling in drop impact:\\ Forces \& maximum spreading diameter}

\author{Vatsal Sanjay}
\email{vatsalsanjay@gmail.com}
\affiliation{	
	Physics of Fluids Group, Max Planck Center Twente for Complex Fluid Dynamics, and J. M. Burgers Center for Fluid Dynamics, University of Twente, P.O. Box 217, 7500AE Enschede, Netherlands
}
\author{Detlef Lohse}
\email{d.lohse@utwente.nl}
\affiliation{	
	Physics of Fluids Group, Max Planck Center Twente for Complex Fluid Dynamics, and J. M. Burgers Center for Fluid Dynamics, University of Twente, P.O. Box 217, 7500AE Enschede, Netherlands
}
\affiliation{
	Max Planck Institute for Dynamics and Self-Organisation, Am Fassberg 17, 37077 G{\"o}ttingen, Germany
}

\date{\today}


\maketitle

\tableofcontents


\section{Direct numerical simulations}\label{sec:NumMethods}

\subsection{Governing equations}

We perform direct numerical simulations of an axisymmetric incompressible drop impacting a rigid surface using the open-source software Basilisk \cite{popinet-basilisk}. The impact dynamics are governed by the droplet's impact velocity $V_0$, diameter $D_0$, and material properties (density $\rho$, kinematic viscosity $\nu$, dynamic viscosity $\mu = \rho\nu$, and surface tension $\gamma$). The conservation of mass and momentum imply

\begin{align}
	\label{eqn:continuity}
	\boldsymbol{\nabla\cdot v} = 0
\end{align}
and

\begin{align}
	\label{Eqn::NS}
	\frac{\partial \boldsymbol{v}}{\partial t} + \boldsymbol{\nabla\cdot}\left(\boldsymbol{v}\boldsymbol{v}\right) &= -\frac{1}{\rho}\boldsymbol{\nabla} p + \boldsymbol{\nabla\cdot}\left(2\nu\boldsymbol{\mathcal{D}}\right),
\end{align}
respectively. Here, $\boldsymbol{v}$ is the velocity field, $p$ is pressure, and $\boldsymbol{\mathcal{D}} = (\nabla\boldsymbol{v} + (\nabla\boldsymbol{v})^T)/2$ is the deformation tensor. For the initial condition, we assume that the drop is about to hit the rigid surface with velocity $V_0$. 
We vary the dimensionless 
control parameters  
$
We \equiv \rho V_0^2D_0 / {\gamma} 
$
and $
Oh \equiv {\mu}/ \sqrt{\rho\gamma D_0}
$
over the huge parameter range $1\le We \le 10^3$ and $10^{-3}\le Oh \le 10^2$. To mimic a liquid-gas free surface, we set the Ohnesorge number of the surrounding gas $Oh_s = (\mu_s/\mu)Oh = 10^{-5}$ and the density ratio $\rho_s/\rho = 10^{-3}$.  
To keep track of this interface, we employ the one-fluid approximation \cite{prosperetti2009computational, tryggvason2011direct} with a volume fraction $\Psi$ to distinguish between the drop ($\Psi=1$) and surrounding air ($\Psi=0$). The volume fraction follows

\begin{align}
	\label{Eqn::Vof2}
	\frac{\partial \Psi}{\partial t} + \boldsymbol{\nabla \cdot}(\boldsymbol{v}\Psi) = 0.
\end{align}

To respect the dynamic boundary condition at the liquid-gas interface, a singular surface tension force $\boldsymbol{f}_\gamma$ is applied at the surface and is given by \cite{brackbill1992continuum}

\begin{equation}\label{Eqn::SurfaceTension}
	\boldsymbol{f}_\gamma \approx \gamma\kappa\boldsymbol{\nabla}\Psi
\end{equation}
where $\gamma$ is the surface tension coefficient and $\kappa$ is the interface curvature calculated using the height-function method. 

The normal force on the substrate is calculated by integrating the pressure over the impact area \cite{landau2013course}:
\begin{equation}
	\label{Eqn::force2}
	\boldsymbol{F}(t) = F(t) \hat{\boldsymbol{z}} = \left(\int_\mathcal{A} \left(p-p_0\right)\mathrm{d}\mathcal{A}\right)\hat{\boldsymbol{z}}
\end{equation}
\noindent where $p_0$ is the ambient pressure, $\hat{\boldsymbol{z}}$ is the unit vector normal to the substrate, and $\mathcal{A}$ is the substrate area. We 
keep track of the maximum spreading diameter $D(t)$ at all times by finding the radial maximum of the spreading drop. 
Another important metric is the viscous dissipation across various regimes for the prediction of the maximal spreading diameter $D_\text{max}$ and the maximal
normal force $F_\text{max}$. The total viscous dissipation per unit mass is $E_d(t) = \int_0^t \varepsilon(t)dt$ where $\varepsilon(t)$ is the viscous dissipation rate per unit time averaged over the drop's mass and is given by

\begin{equation}
	\varepsilon(t) = \frac{3}{4\pi \rho D_0^3}\int_m\zeta_\nu(t, \boldsymbol{x}) dm,
	\label{Eqn::dissipation1}
\end{equation}
where, $\zeta_\nu(t, \boldsymbol{x}) = 2\nu \left(\boldsymbol{\mathcal{D}:\mathcal{D}}\right)$  is the local viscous dissipation function \cite{landau2013course}. Both $\varepsilon(t)$ and $\zeta_\nu(t, \boldsymbol{x})$ have the dimensions of $V_0^3/D_0$, i.e., length squared over time cubed or velocity squared over time, as it should be for dissipation rate of energy per unit mass.

\begin{figure}
	\centering
	\includegraphics[width=\textwidth]{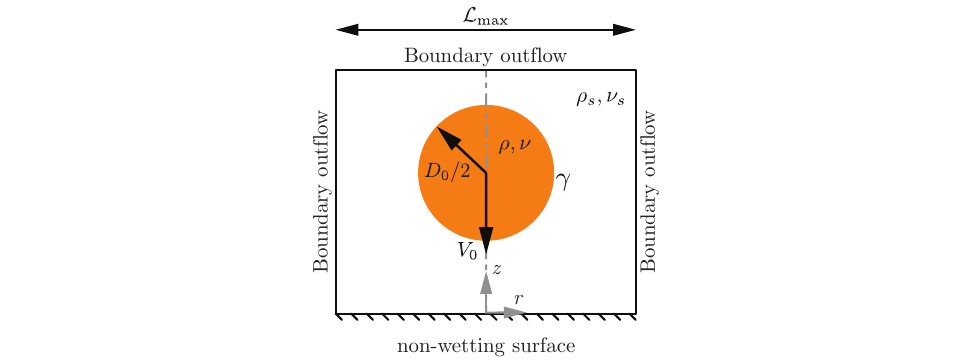}
	\caption{Problem schematic with an axisymmetric computational domain used to study the impact of a drop with diameter $D_0$ and velocity $V_0$ on a non-wetting substrate. The grey dashed-dotted line represents the axis of symmetry, $r = 0$. Boundary air outflow is applied at the top and side boundaries (tangential stresses, normal velocity gradient, and ambient pressure are set to zero). The domain boundaries are far enough from the drop not to influence its impact process ($\mathcal{L}_\text{max} \gg D_0$, $\mathcal{L}_\text{max} = 8R$ in the worst case).}
	\label{fig:S1}
\end{figure}

\subsection{Simulation details}
\subsubsection{Domain description}
We leverage the axisymmetry of the drop impact process to reduce the computational complexity. Fig.~\ref{fig:S1} illustrates the axisymmetric domain used in our simulations, where $r=0$ denotes the axis of symmetry. The domain size is chosen to be sufficiently large ($\mathcal{L}_\text{max} \gg D_0$, with $\mathcal{L}_\text{max} = 8D_0$ in the worst case) to avoid boundary effects on the impact dynamics.
At the substrate ($y=0$), we impose no-slip and non-penetration conditions for velocity, along with a zero normal gradient condition for pressure. To model a perfectly non-wetting surface, we maintain a thin air layer between the drop and the substrate by setting the VoF tracer $\Psi = 0$ at the wall. This approach, while not fully resolving microscopic dynamics within the air layer, has been shown to accurately capture the macroscopic behavior of drop impact on superhydrophobic surfaces \citep{ramirez2020lifting,sanjay2023drop,alventosa2023inertio,garcia2024skating}.
At the top and side boundaries, we apply outflow conditions, setting tangential stresses and normal velocity gradients to zero, and pressure to the ambient value. These boundary conditions allow for the unimpeded exit of air from the domain during impact.
\subsubsection{Brief note on the discretization schemes}
We employ a finite volume discretization on a staggered grid, with second-order accuracy in both space and time \cite{popinet-basilisk}. The momentum advection terms are treated using the conservative scheme \cite{popinet-basilisk-momentum}, while viscous terms are handled implicitly \cite{popinet-basilisk-viscosity}. Surface tension is implemented using the well-balanced continuum surface force method \cite{popinet2018numerical,basiliskPopinet2}.
Adaptive mesh refinement is utilized to efficiently resolve regions of high velocity gradients and the drop-air interface. The minimum grid size is set to $\Delta = D_0/2048$, which we determined through mesh independence studies to be sufficient for accurate resolution of the impact dynamics.
The maximum time step is restricted by the capillary wave stability criterion to ensure stability of the explicit surface tension scheme \citep{popinet2009accurate,basiliskPopinet2}. For the VoF advection, we use a geometric method that ensures mass conservation and sharp interface representation.

\subsubsection{Initial conditions}
At $t=0$, we initialize a spherical drop with its south pole positioned $0.05D_0$ above the substrate, falling with velocity $V_0$. While this idealized initial condition may not fully capture the residual oscillations present in experimental drops, particularly for low $We$ and $Oh$, we expect that the influence of these shape variations is at least partially accounted for in the experimental error bars derived from repeated trials.

\subsubsection{Limitations and considerations}
It is worth noting that the axisymmetric assumption breaks down for high Weber numbers ($We \gtrsim 100$ for water drops, and even higher for more viscous fluids) due to destabilization by the surrounding gas after splashing \citep{xu2005drop,Eggers2010,Driscoll2011,riboux2014experiments,Josserand2016,sanjayzhang2022prl}. While greatly influencing the second peak amplitude of the impacting force, the maximal force right after impact is invariant to this change in dynamics \cite{sanjayzhang2022prl,sanjay2024inertia}. For the maximum spreading diameter, we disregard any smaller droplets that might be ejected from post lamella destabilization of the impacting drop. 
Furthermore, while our thin air layer approach effectively models an ideal non-wetting surface, it does not fully resolve the microscopic dynamics within the air film \cite{kolinski-2014-epl,sprittles2024gas}, particularly when it thins below a critical size of approximately $10\Delta$. However, for the parameter range of interest in this study, this approach has been demonstrated to accurately capture the macroscopic impact behavior \citep{ramirez2020lifting,sanjay2023drop,alventosa2023inertio,garcia2024skating}.
The simulation source codes and post-processing scripts used in this study are available in a public GitHub repository \citep{basiliskVatsal}, ensuring reproducibility and facilitating further development of this work.

\section{Regression procedure for finding the maximum impact force}

The expression for the maximum impact force is given by (eq.~(12) of the original text):

\begin{align}
	\label{supp:eq:FinalFmax}
	\frac{F_{\text{max}}}{\rho V_0^2D_0^2} =\, \beta_0 + \frac{\beta_1}{We} + \left\{ \begin{aligned} 
		&k_0 We^{-1/4}\sqrt{Oh} && \text{for}\,Re > 1\\ 
		&m_0 + m_1\left(Oh/\sqrt{We} - 1\right) && \text{for}\,Re < 1 
	\end{aligned} \right.
\end{align}

To determine the coefficients in this equation, we employ a multi-step regression procedure. We first consider the inertial limit where we have previously \cite{sanjayzhang2022prl,sanjay2024inertia} shown that $F_{\text{max}} \approx 0.81 + 1.6/We$ over two orders of magnitude in $Oh$. This allows us to set $\alpha_0 = 0.81$ and $\alpha_1 = 1.6$. To account for viscous effects, we focus on regions far from the transition line between inertial and viscous regimes. Specifically, we consider the data points with $Re \geq 10$ for the inertial regime and $Re \leq 0.1$ for the viscous regime. We perform separate fits to minimize the least-squared errors in these regions to obtain initial estimates for $k_0$, $m_0$, and $m_1$. To model the transition between regimes, we introduce a smooth transition function using a {\it Sigmoid} function centered at $Re = 1$, given by
\begin{align}
	\sigma(Re) = \frac{1}{1 + \exp(-\chi\left(Re - 1\right))}
\end{align}
where $\chi$ is the smoothness parameter for the {\it Sigmoid}. This ensures a continuous transition between the inertial and viscous expressions. Finally, we perform a global least-squared fit using all data points to find the value of $\chi$ and refine the values of $k_0$, $m_0$, and $m_1$, while keeping $\alpha_0$ and $\alpha_1$ fixed at their inertial limit values. This procedure results in the following expression for the maximum impact force:
\begin{align}
	\label{FinalFmaxWithCoeff}
	\frac{F_{\text{max}}}{\rho V_0^2D_0^2} =\, 0.81 + \frac{1.6}{We} + \left\{ \begin{aligned} 
		&1.44 We^{-1/4}\sqrt{Oh} && \text{for}\,Re > 1\\ 
		&3.85 + 0.72\left(Oh/\sqrt{We} - 1\right) && \text{for}\,Re < 1 
	\end{aligned} \right.
\end{align}
with the fit being fairly insensitive to $0.8 < \chi < 1.2$.

\section{Regression procedure for finding the maximum spreading diameter}
\begin{figure}
	\centering
	\includegraphics[width=\textwidth]{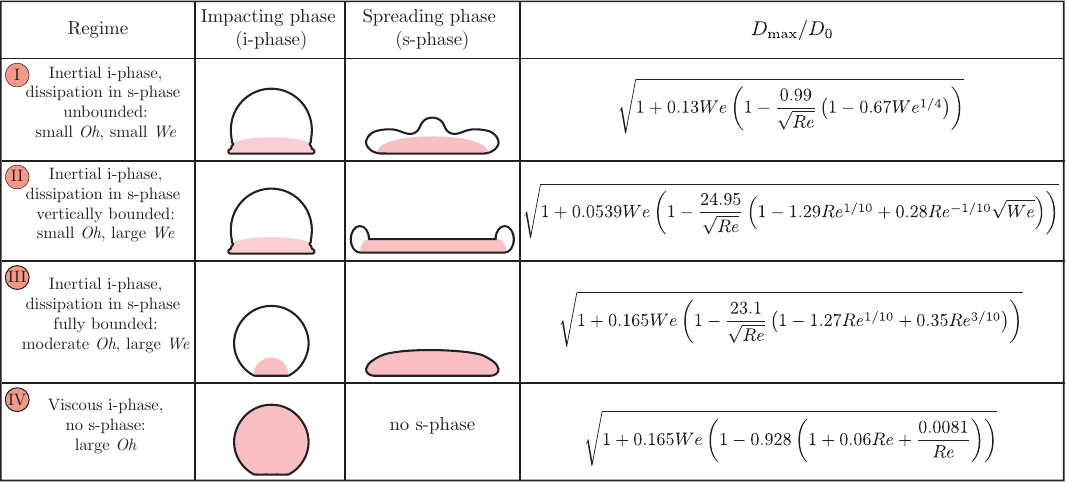}
	\caption{Relations to predict the maximum spreading diameter for the four regimes I, II, III, and IV.  }
	\label{supp:dmax_results}
\end{figure}

The expression for the maximum spreading diameter is given by (eq.~(4) of the original text):
\begin{align}
	\label{suppleq:GlobalDmax_v2}
	D_{\text{max}} &= D_0\sqrt{\left(\alpha_0 + \alpha_1We\left(1 - \frac{E_d(t = \tau_s)}{V_0^2}\right)\right)}. 
\end{align}
where the dimensionless dissipation (eq.~(16) of the original text) is given by
\begin{align}
	\label{suppleq:FinalDmax}
	\frac{E_d(t = \tau_s)}{V_0^2} \sim \begin{cases}
		\left(a_0 + a_1We^{1/4}\right)/\sqrt{Re}\quad\text{in Regime I,}\\
		\left(b_0 + b_1Re^{1/10} + b_2Re^{-1/10}We^{1/2} \right)/\sqrt{Re}\quad\text{in Regime II,}\\
		\left(c_0 + c_1Re^{1/10} + c_2Re^{3/10}\right)/\sqrt{Re}\quad\text{in Regime III,}\\
		d_0 + d_1Re + d_1/Re\quad\text{in Regime IV,}
	\end{cases}
\end{align}

To determine the coefficients, we employ a multi-step regression procedure. Initially, we fit the scaling laws in each regime separately, focusing on data points far from transition lines to obtain preliminary estimates for $\alpha_i$, $a_i$, $b_i$, $c_i$, and $d_i$. To model regime transitions, we introduce smooth transition functions using Sigmoid functions. We define Euclidean distance functions for each transition:
\begin{alignat}{2}
	R_\text{I--II}   &= \sqrt{(Oh - Oh_{c,\text{I--II}})^2 + (We - We)^2},   &&\text{ where } Oh_{c,\text{I--II}} = We^{-2} \\
	R_\text{II--III} &= \sqrt{(Oh - Oh_{c,\text{II--III}})^2 + (We - We)^2}, &&\text{ where } Oh_{c,\text{II--III}} = We^{-3/4} \\
	R_\text{III--IV} &= \sqrt{(Oh - Oh_{c,\text{III--IV}})^2 + (We - We)^2}, &&\text{ where } Oh_{c,\text{III--IV}} = We^{1/2}
\end{alignat}
Using these distances, we define Sigmoid transition functions:
\begin{align}
	\sigma_i(R_i) = \frac{1}{1 + \exp(-\chi_i R_i)}
\end{align}
where $i$ represents each transition and $\chi_i$ are smoothness parameters.
We then combine individual regime expressions using these transition functions to create a global model:
\begin{align}
	\frac{D_{\text{max}}}{D_0} = (1-\sigma_\text{I--II})f_\text{I} +
	\sigma_\text{I--II}(1-\sigma_\text{II--III})f_\text{II} +
	\sigma_\text{II--III}(1-\sigma_\text{III--IV})f_\text{III} +
	\sigma_\text{III--IV}f_\text{IV}
\end{align}
where $f_\text{I}$, $f_\text{II}$, $f_\text{III}$, and $f_\text{IV}$ are the expressions for each regime after combining eqs.~\eqref{suppleq:GlobalDmax_v2}--\eqref{suppleq:FinalDmax}.
Finally, we perform a global least-squared fit using all data points to refine all parameters by minimizing the mean squared error between model predictions and numerical simulation results. This procedure yields expressions for the maximum spreading diameter in each regime, as shown in fig.~\ref{supp:dmax_results}. The global model provides smooth transitions between regimes and accurately captures behavior across the entire parameter space. We find the fit to be relatively insensitive to smoothness parameters in the range $0.8 < \chi_i < 1.2$.

	
\bibliographystyle{prsty_withtitle}
\bibliography{refs}